RESEARCH ARTICLE

RUNNING HEAD: CELL-CELL CONTACT MEDIATES PHENOTYPE TRANSITIONS

# Intercellular contact is sufficient to drive fibroblast-to-myofibroblast transitions

Vasuretha Chandar[1], Benjamin M. Goykadosh[1] & Harikrishnan Parameswaran[1]

[1]Department of Bioengineering, Northeastern University, Boston, MA, USA

Correspondence: *Harikrishnan Parameswaran, Ph.D. (h.parameswaran@northeastern.edu)*

## ABSTRACT

Fibroblast cells play a key role in maintaining the extracellular matrix. During wound healing, fibroblasts differentiate into highly contractile myofibroblasts, which secrete extracellular matrix proteins like collagen to facilitate tissue repair. Under normal conditions, myofibroblasts undergo programmed cell death after healing to prevent excessive scar formation. However, in diseases like fibrosis, the myofibroblasts remain active even after the wound is closed, resulting in excessive collagen buildup and a stiff, fibrotic matrix. The reasons for the persistence of myofibroblasts in fibrosis are not well understood. Here, we show the existence of a mechanism where direct physical contact between a fibroblast and a myofibroblast is sufficient for fibroblasts to transition into myofibroblasts. We demonstrate that the fibroblast-myofibroblast transition can occur even in the absence of known biochemical cues, such as growth factor activation or mechanical cues from a stiff, fibrotic matrix. Furthermore, we demonstrate that contact-based fibroblast-myofibroblast activation can be inhibited by the Gαq/11/14 inhibitor FR900359, which prevents the formation of myofibroblasts. These findings provide new insights into the persistence of fibrosis despite therapeutic interventions, suggesting a potential strategy for targeting the fibroblast-to-myofibroblast transition in fibrotic conditions.

## NEW & NOTEWORTHY

This study uncovers a novel mechanism of fibroblast-to-myofibroblast transition (FMT) driven by cell-cell contact. Myofibroblast-fibroblast contact elevates cytoskeletal tension in the fibroblast, which in turn drives transition via activation of the GqGPCR signaling pathway. This process occurs independently of exogenous biochemical cues or stiff matrices. Emphasizing the central role of cytoskeletal stress in FMT, these findings reveal a mechano-regulated pathway, opening new avenues for therapeutic intervention in fibrotic diseases.

---

## INTRODUCTION

Scar formation is a common pathological outcome of chronic injury across organs, reflecting a dysfunctional extension of normal wound-healing programs. At the site of injury, one of the key cellular processes is the transformation of resident fibroblasts into highly contractile myofibroblasts(1). These specialized cells, characterized by α-smooth muscle actin (α-SMA) expression, play a crucial role in wound contraction and tissue remodeling by depositing extracellular matrix components, particularly collagen, to facilitate tissue repair (2–8). In normal wound healing, once tissue integrity is restored, myofibroblasts are typically deactivated and cleared from the site(9,10). However, in diseases such as asthma, where fibrotic airway remodeling occurs or idiopathic pulmonary fibrosis, this repair program becomes dysregulated, leading to persistent myofibroblast activation and accumulation(10–13). This results in excessive collagen deposition, causing scarring, stiffening, and fibrosis of the lung tissue. The stiffening of the lung severely impairs the ability of the lungs to facilitate oxygen and carbon dioxide exchange. Thereby causing organ dysfunction, leading to hypoxemia, loss of quality of life, and ultimately death(14,15). As such, the persistence of the myofibroblast phenotype plays a crucial role in the progression of fibrosis.

Current therapeutic strategies aim to prevent or reverse myofibroblast activation through two main approaches. The first involves targeting canonical biochemical pathways and growth factors such as the WNT/β-catenin, PDGF, TGFβ, and TNF-α among others(16). Existing drugs (ex: Nintedanib, Pirfenidone) inhibit growth factors and cytokines, which prevent the release of inflammatory signals that trigger the transition of fibroblasts to myofibroblasts(17,18). The second approach recognizes the mechanosensitive nature of fibroblasts and focuses on matrix remodeling. Fibroblasts can sense the stiff fibrotic environment created by excessive collagen deposition and transform into myofibroblasts, that secrete more collagen reinforcing a detrimental feedback loop(19–21). To counteract this cycle, ongoing in vitro and in vivo studies are investigating matrix remodeling as a potential therapeutic approach(22–24). These studies utilize collagen-degrading enzymes, such as matrix metalloproteinases (MMPs), to soften the matrix and potentially halt the disease progression(25–27). However, despite these interventions, the myofibroblast phenotype continues to persist, suggesting additional mechanisms are at play that remain unknown.

The goal of this study is to identify mechanisms that explain why myofibroblast-like behavior can persist in disease even when canonical exogenous pro-fibrotic signals such as a stiff matrix or TGFβ are limited.

In this work, we follow a stepwise logic from exclusion to mechanism. By isolating cell-cell contact from these inputs, we identify a previously underappreciated mechanism for sustained myofibroblast activation. Using independent, functionally relevant readouts of myofibroblast markers, cytoskeletal tension, and calcium signaling, we show that direct cell-cell mechanical interactions are sufficient to drive fibroblast to myofibroblast transitions.

## MATERIALS AND METHODS

1. **Fabrication of an optically clear substrate:** NuSil is an optically clear, biologically inert PDMS substrate with a tunable Young's modulus ranging from 0.3 to 70kPa. To model soft healthy tissue, equal parts of NuSil gel-8100, A and B (NuSil, CA, USA) without any crosslinker (0.0%) were mixed (1:1) to obtain a substrate with Young's modulus E = 300Pa(28), similarly, a crosslinker volume 0.07% of the combined parts A and B volume was added to obtain soft substrates with Young's modulus E = 1.9kPa(29) and 0.36% to obtain stiff substrate E = 13kPa(30) Each mixture was spin-coated onto 30-mm-diameter, #1.5 glass coverslips for 50 seconds to produce a 100µm-thick layer of substrate. Coverslips were incubated at room temperature for an hour and cured overnight at 60°C. These cured coverslips were then secured in sterile 40mm Bioptech dishes (Biological Optical Technologies, PA, USA), UV sterilized for 1 hour, and protein coated with 0.1% gelatin solution (Millipore Sigma, Cat #ES-006-B) for use in cell culture experiments. Substrate stiffness was measured in accordance with the method outlined by Yoshie et al (31). A Discovery HR-3 Peltier Plate Aluminium Rheometer with a 40mm parallel plate (serial #5333-0647) was utilized. 40mm coverslips coated with Nusil gel prepared with varying crosslinker volumes, were measured by applying a strain of 0.5% at a frequency of 1Hz. The stiffness of the substrate is converted to Youngs' modulus (E) using the method published and described in detail(31).

2. **Cell Culture:** NIH 3T3 mouse fibroblasts (CRL-1658, Lot #70030679) acquired from ATCC were utilized for all our experiments. Primary healthy normal human lung fibroblasts (NHLFs; Lonza, CC-2512; donor: 3-year-old Caucasian male) were used throughout this study wherever indicated. Cells were grown under standard culture conditions of 37°C and 5% CO2 and used by passage 20 mouse fibroblasts and by passage 8 for primary cell lines for all experiments. Culture medium: Cells were cultured in 10% fetal bovine serum, Dulbecco's modified Eagle's medium/F12 (Fisher Scientific, Massachusetts), 1X penicillin/streptomycin (Fisher Scientific), 1X MEM nonessential amino acid solution (Sigma Aldrich, Missouri), and amphotericin B (25 µg/liter; Sigma-Aldrich). Co-culture platform to study fibroblast-to-myofibroblast transition: Equal number of fibroblasts and myofibroblasts (1:1) were plated together on the 300Pa soft substrate and cultured atleast for 72 hours, unless otherwise stated. For co-culture experiments, cells were seeded at a density of $10^5$ cells/cm$^2$. For experiments with a physical barrier on 300Pa, an optimized density of 150,000 cells of each cell type were added per side of the divider.

3. **Generation of a fibroblast and myofibroblast population:** To visualize and distinctively identify fibroblasts, NIH 3T3-mouse fibroblasts (CRL-1658, Lot #70030679) or Primary healthy normal human lung fibroblasts (NHLFs; Lonza, CC-2512) were transduced with a lentiviral vector encoding tandem-dimer Tomato fluorescence protein (tdTomato, Vector Builder, catalog no. VB181014-1005thm). Briefly, the cells were seeded and cultured according to standard cell culture protocol and after 18 hours, cells were introduced to cell culture medium containing polybrene (5 µg/ml; Thermo Fisher Scientific, catalog no. TR1003G) and the manufacturer's recommended functional titer concentration of lentiviruses. After 24 hours of exposure to the transduction medium, cells were allowed to recover in unadulterated cell culture medium for an additional 48 hours. Next, cells were harvested using

Gibco TrypLE Select reagent and cell selection was performed by culturing cells on 300Pa soft substrate in culture medium containing blasticidin (8 µg/ml; Sigma-Aldrich, catalog no. 15205). This 3T3-mouse fibroblast population expressing tdTomato was expanded on protein coated 300Pa soft substrates and cultured for up to 6 weeks before being used in experiments and used within 20 passages since the time of thawing. To generate a myofibroblast population, non-transduced 3T3-mouse fibroblasts were cultured for a period of at least 4 weeks on tissue culture plastic (≥100kPa) and utilized within passage 20 for all experiments. Myofibroblasts were identified as cells that stained positive for α-SMA in stress fibers. Prior to utilizing the two cell populations for experiments, validation was performed by assessing αSMA localization to confirm cell phenotypes: fibroblasts on soft substrates consistently lacked αSMA-positive stress fibers, whereas myofibroblasts on stiff substrates exhibited robust αSMA incorporation into stress fibers.

4. **Immunostaining and microscopy:** Cells were fixed in 4% paraformaldehyde in PBS, Thermo Scientific Chemicals (Cat #J61899.AK) at room temperature for 10 minutes and washed 3 times with 1X PBS, Gibco (Cat # 14040133). Cells were then permeabilized with 100% ethanol, Thermo Scientific (Cat # T038181000) or 0.3% Triton-X, Thermo Scientific Chemicals (Cat # A16046.AE) for 10 min at room temperature. Following permeabilization, cells were blocked using 1X PBS containing 0.1% Tween-20, 1% bovine serum albumin (BSA) Thermo Scientific (Cat # 37520) for 30 min at room temperature. Next, cells were labelled with Alpha-Smooth Muscle Actin Monoclonal Antibody (1A4), eFluor™ 660, eBioscience™ (Cat #50-9760-82) at 5 µg/mL in 1X PBS containing 1% BSA and simultaneously stained for Filamentous actin with ActinGreen™ 488 ReadyProbes™ Reagent according to manufacturer protocol (AlexaFluor™ 488 phalloidin, Cat# R37110) or with YAP (D8H1X) Rabbit Monoclonal

Antibody at 1:200 dilution in blocking solution (Alexa Fluor® 488 Conjugate #14729) for 1 hour at room temperature. Cells were washed 3 times with 1X PBS to remove non-specific binding. Lastly, cell nuclei were labeled with NucBlue (Fisher Scientific, Waltham, MA, USA) and mounted using ProLong™ Glass Antifade Mountant (Cat #P36984) according to manufacturer protocol. Images were acquired with a Leica DMi8 inverted microscope, a Leica DFC6000 camera, and a Leica LED8 light engine (excitation, 480/40 nm; emission, 527/50 nm) using a 20X/0.55 dry objective or 40X/0.65 dry objective (Leica DMi8 microscope, Leica, Wetzlar, Germany).

5. **Applying a point displacement on the matrix:** We fixed a nail dotting tool (ORLY, California) with a 1.5 mm diameter ball-end to the InjectMan 4 micromanipulator (Eppendorf, Germany) on our Leica DMi8 microscope. The micromanipulator allowed precise control of this tool with which we applied a transient poke to the extracellular matrix some distance away (300µm) from adhered cells at a 45˚ angle and depth of 30µm to generate a shear stress of the order of 10s of Pa, comparable to that of a myofibroblast.

6. **Fluorescent PIP2 and Calcium ($Ca^{2+}$) Imaging:** Fibroblasts were loaded with a fluorescent cytosolic PIP2 indicator to record changes in [PIP2] using a live green downward phosphatidylinositol 4,5-biphosphate (PIP2) assay kit (#D0400G), Montana Molecular (Montana). The assay genetically encodes fluorescent protein expression using a BacMam vector and fluorescence intensity decreases with increasing PIP2 hydrolysis. Following the manufacturer protocol provided, we transduced cultured fibroblasts with the PIP2 sensor for 24 hours at 37°C and 5% $CO_2$. Prior to imaging, cells were washed with 1X Hanks' Balanced Salt Solution (HBSS) and incubated at 37°C and 5% $CO_2$ for an additional 30 minutes. The cells were washed once more before imaging. To record changes in calcium concentration [$Ca^{2+}$], cells were loaded with a fluorescent cytosolic $Ca^{2+}$ indicator, FLIPR $Ca^{2+}$ 6 (Molecular Devices, California). Fluorescence intensity increases with increasing cytosolic [$Ca^{2+}$] and decreases as [$Ca^{2+}$] declines. FLIPR $Ca^{2+}$ 6 was prepared according to the manufacturer's standards. Cells were incubated with a 1:1 solution of FLIPR $Ca^{2+}$ 6 and cell culture medium at 37°C and 5% CO2 for 30 minutes before imaging. 16-bit images were recorded at 0.5 Hz for 5 minutes before indentation (~20Pa) and at least 5 minutes post indentation. Cells were imaged with a Leica DMi8 inverted microscope, a Leica DFC6000 camera, and a Leica LED8 light engine (excitation, 480/40 nm; emission, 527/50 nm) with a 40X/0.65 dry objective. The excitation and emission wavelengths for the PIP2 was ex:480 nm/ em:515 nm. To analyze data, each image sequence was loaded in Fiji ImageJ, and regions of interest (ROIs) were hand-selected in the cytoplasm of each cell to obtain mean grayscale intensities over the area of the ROI for each frame in time. Obtained intensities were normalized with the baseline average to plot normalized intensity values using GraphPad Prism ver. 10.4.1.

7. **Patterning cell ensembles:** We utilized the Alvéole's PRIMO optical module and the Leonardo software, a UV-based, contactless photopatterning system (Alvéole, Paris, France). 13kPa NuSil substrates coated with a layer of fluorescent beads as fiducial markers for traction force microscopy were utilized. A 5% solution of 0.2-µm-diameter green, fluorescent carboxylate-modified microspheres (FluoSpheres, Invitrogen, Carlsbad, CA, USA) in PBS was vortexed for 10 seconds. Solution (2 ml) was added to each substrate in a Bioptech dish and left at room temperature for 1 hour to allow the beads to adhere. The bead solution was poured off, substrates were washed three times with PBS, and then PBS was poured off. Substrates were then protein coated with 0.1% gelatin solution for use in cell culture. The substrate surface was first coated with 500 µg/mL PLL (Sigma Aldrich, St. Louis, MO, USA) for 1 hour at room temperature. The substrate was washed with PBS and 10 mM HEPES buffer adjusted to pH 8.0 and incubated with 50 mg/mL mPEG-SVA (Laysan Bio, Inc., Arab, AL, USA) at room temperature for 1 hour, and washed with PBS once more. The PRIMO system was calibrated using fluorescent highlighter on an identical substrate. The PBS was replaced by 14.5 mg/mL PLPP (Alvéole, Paris, France), and then the desired pattern, previously created with graphic software, was illuminated with UV light focused on the substrate surface for 30 seconds. Patterned surfaces were washed again with PBS and then incubated with 0.1% gelatin for 1 hour at room temperature. The substrate was washed and hydrated in standard cell culture media. Fibroblasts and myofibroblasts were added in a 1:1 ratio totaling a cell density of $3X10^4$ cells/ml to obtain a confluent pattern of the two cell types within 24 hours.

8. **<u>Traction Force Microscopy and Monolayer Stress Microscopy measurements:</u>** Before the experiment, cell culture media was replaced with $Ca^{2+}/Mg^{2+}$ free HBSS and cells were incubated at 37 °C for 30 minutes. Tractions were recorded by imaging the fluorescent beads with a 20X/0.55 dry objective and the Leica DMi8 microscope. Continuous time-lapse imaging was performed at baseline and throughout the experimental period. Images of beads were taken after cells were removed (using RLT Lysis Buffer, Qiagen, Hilden, Germany). Using these images, cellular forces were calculated with a custom MATLAB (MathWorks, Natick, MA, USA) software program using Fourier Traction Force Microscopy (TFM). To calculate the net contractility of each cell within a multicellular ensemble, we employed monolayer stress microscopy (MSM). This computational approach estimates stresses in a cell layer based on the forces at the cell-matrix interface. Initially, traction force vectors obtained from TFM were processed using a modified version of the open-source Python code, pyTFM(32). By applying a mask that outlines cell-cell borders from a phase-contrast image, we were able to separately calculate contractility due to cell-cell interactions, cell-matrix interactions, and the net contractility of each cell, as well as determine the normal and shear forces at each cell border.

9. **Measurement of nuclear spread area:** The nuclear spread area represented in this manuscript is the average nuclear area of 10 cells across 10 randomly selected non-overlapping regions in each dish (N). 40X images were loaded into Fiji ImageJ, and regions of interest (ROIs) were hand-selected in the DAPI images using the ellipse tool, by tracing the outline of the nucleus. The area of selection was spatially calibrated and represented as $\mu m^2$. YAP staining intensity in the nucleus and within the cytoplasm was quantified using the software ImageJ (National Institutes of Health, Bethesda, MD); these intensities were used to calculate a nuclear/cytoplasmic YAP.

10. **Data representation and statistical testing:** Throughout the manuscript, we represent data as the average of N independent trials (biological replicates). For each dish, images were acquired at 40X magnification capturing at least 109 ± 18 cells per field of view. 10 non-overlapping square regions (0.34 X 0.34 mm each) were randomly selected (technical replicates) for each biological replicate to ensure comprehensive sampling of the substrate. Unless otherwise stated, all data correspond to representative images collected at 40X. The error bars indicate SD over independent trials. The individual data points are shown in all bar plots. SigmaPlot ver.11 build 11.0.0.75 (Systat Software, San Jose, CA) and GraphPad Prism ver. 10.4.1. were used to perform statistical tests and calculate effect sizes ($\eta^2$ or $\eta p^2$). One-way ANOVAs followed by post-hoc pairwise comparisons were used to test significant differences in datasets of three or more groups, which were influenced by one independent factor. Pairwise comparisons were performed using the t-test when data were normally distributed. Otherwise, the Mann-Whitney rank-sum test was used to compare median values. Statistical significance was reported using asterisks, where $p < 0.05$ (*), $p < 0.01$ (**), $p < 0.001$ (***).

The specific tests used, number of samples, $\eta^2$ and p values are described along with the corresponding results.

# RESULTS

## 1. Fibroblasts can transition into myofibroblasts even in the absence of exogenous biochemical signals or mechanical cues from a stiff extracellular matrix.

To test whether fibroblasts can transition to myofibroblasts without external chemical or mechanical cues, we generated two distinct cell populations. First, a fibroblast population that does not express α-SMA in stress fibers was generated by culture on a 300Pa soft substrate for 6 weeks. Similarly, a myofibroblast population expressing α-SMA in stress fibers was generated by culturing fibroblasts on tissue culture plastic (E$\geq$100kPa), for 4 weeks. To identify fibroblasts in a co-culture with myofibroblasts, we labeled fibroblasts with a fluorescent tdTomato reporter (magenta). Myofibroblasts that express the reporter indicate the fibroblast that transitioned into a myofibroblast in culture. The cells were then seeded to maintain a 1:1 ratio of fibroblasts to myofibroblasts ($10^5$ cells/$cm^2$) on a 300Pa (soft substrate) for 72 hours under standard cell culture conditions. No external mechanical signals (Stiff substrate) or chemical stimuli (e.g., TGFβ) were introduced to facilitate phenotype transitions. After 72 hours, the fibroblast cells were stained for filamentous actin (F-actin) and alpha-smooth muscle actin (α-SMA) in the stress fibers. To assess their phenotypic characteristics, we also measured the nuclear spread area of fibroblasts and myofibroblasts(33). We counted the number of transformed fibroblasts in each field of view (40X) and found that 20$\pm$5 fibroblasts in the co-culture expressed αSMA in their stress fibers as compared to only 1.66$\pm$0.57 cells in fibroblast mono-cultures (**Fig. 1A-C,** p=0.0032, t-test, N=3). Further, the was no significant difference in the nuclear spread area of fibroblasts in a co-culture compared to myofibroblasts in the co-culture, confirming a phenotypic transition of fibroblasts into myofibroblasts (**Fig. 1D-E**, One-way ANOVA, N=3). We also assessed YAP subcellular localization during fibroblast-myofibroblast contact. Our findings demonstrate increased YAP nuclear translocation in fibroblasts that are in contact with myofibroblasts in comparison to fibroblasts in mono-culture (see **Supplemental Figure S1**). To enhance translational relevance, co-culture experiments were repeated using primary normal human lung fibroblasts (see **Supplemental Figure S2**). Notably, fibroblast transition to myofibroblast was observed in both co-cultures of immortalized and primary cell lines in the absence of external chemical or mechanical stimuli. This suggests that cell-cell interactions within the co-culture environment may be responsible for driving this phenotypic change.

### 2. Intercellular contact facilitates fibroblast-to-myofibroblast transition (FMT)

To better understand the mechanism underlying the transition of fibroblasts to myofibroblasts, even in the absence of external cues, we hypothesized that fibroblasts and myofibroblasts communicate with each other to facilitate FMT. To test this hypothesis and identify the primary mechanism for FMT transitions in co-culture, we used a 3D-printed barrier (~1.5mm thick) to split our culture dishes into two halves. (**Fig. 2A**) On one side of the barrier, we cultured myofibroblasts, and on the other side we cultured fibroblasts. We maintained the 1:1 ratio of fibroblasts and myofibroblasts (as in **Fig. 1**). On both sides of the barrier, cells were cultured on soft (E=300Pa) NuSil gels. The barrier prevented contact between fibroblasts and the myofibroblasts. To assess barrier permeability, we introduced 1 µm-diameter red fluorescent beads to one side of the divider and imaged diffusion at 1 Hz for 30 s. The rapid, smooth spread of these beads across the dish confirmed that the barrier permits diffusion of any soluble signals as large as 1 µm (**see Supplemental Video S1**). Cells were cultured under standard cell culture conditions for 72 hours. The cells were then stained for f-actin and αSMA in the stress fibers to determine whether fibroblasts transformed into myofibroblasts. Cells that underwent FMT were identified by the presence of tdTomato reporter and alpha-smooth muscle actin in the stress fibers (**Fig. 2C**). We counted the number of fibroblasts that underwent FMT within a field of view, averaged over N=3 independent trials. With the barrier in place preventing direct physical contact, only a statistically insignificant number of fibroblasts (1.3±0.57 cells, p<0.001, unpaired t-test) showed signs of alpha-smooth muscle actin in stress fibers. However, after removing the barrier, 20±2 fibroblasts transitioned into myofibroblasts (**Fig. 2D**). These findings indicate that cell-to-cell contact between a fibroblast and a myofibroblast is sufficient for the fibroblast-to-myofibroblast transition.

### 3. Contact between fibroblasts and myofibroblasts results in an increase of cytoskeletal tension in fibroblasts

As a next step, we investigated the potential role of mechanosensitive signaling in the cell-contact-dependent transition of fibroblasts to myofibroblasts. Prior research has demonstrated that fibroblasts exhibit lower intrinsic cytoskeletal tension (cellular pre-stress) compared to myofibroblasts (22,34). We hypothesized that upon contact with myofibroblasts, the high cytoskeletal tension in myofibroblasts may alter the cytoskeletal tension in fibroblasts, thereby triggering the fibroblast-to-myofibroblast transition. In our study, we used a publicly available implementation of Monolayer Stress Microscopy (MSM) to quantify the mechanical stresses within the cells in the co-culture and calculate the net cellular contractility ($M_{ii}$)(32,35). The MSM technique combines Traction Force Microscopy (TFM) with computational modeling to generate stress maps, allowing us to visualize and measure the internal mechanical tension across the cell monolayer(36–38). Briefly, the fibroblasts and myofibroblasts were co-cultured into a multicellular photopatterned ensemble (**Fig. 3A**). Cells were imaged after 24 hours of seeding, and we measured the traction forces exerted by cells onto a deformable E=13kPa substrate embedded with fluorescent beads. By applying a mask that outlines cell-cell borders from a phase-contrast image, we reconstructed the force vectors at the cell-substrate interface (**Fig. 3B**). Using these traction forces as input, MSM enabled us to compute the in-plane stresses within the monolayer, providing a detailed map of mechanical tensions between cells where we were able to separately calculate contractility due to cell-cell interactions, cell-matrix interactions, and the net contractility of each cell, as well as determine the normal and shear forces at each cell border by analyzing bead displacements (32). The underlying assumptions and limitations of this method have been detailed in previous work (35).

We observed that fibroblasts in direct contact with myofibroblasts in the co-culture exhibited increased cytoskeletal tension ($M_{ii}$), showing no statistical difference between myofibroblast and the first neighbor fibroblasts on day 1 (**Fig. 3C,** Mann-Whitney Rank Sum Test, p=0.004, N=2). This data is particularly of interest because no visible phenotypic changes are detected in the fibroblasts after 24 hours of co-culture with myofibroblasts on 300Pa soft substrate. Specifically, we do not observe the expression of αSMA in stress fibers (**Fig. 3D**) and the nuclear spread area of fibroblasts is significantly lower than that of myofibroblasts (**Fig. 3E**, One-way ANOVA, p=0.034, N=3). The findings from this work are the first to suggest that an increase in cytoskeletal tension precedes phenotypic transition from fibroblasts to myofibroblasts, providing insight into the mechanical aspects of this cellular transformation.

### 4. An increase in cytoskeletal tension is sufficient to drive the fibroblast-to-myofibroblast transition

To determine if an increase in cell prestress alone is sufficient to drive fibroblast-to-myofibroblast transition (FMT), fibroblast mono-cultures on the 300Pa substrate were treated with 2.5μM histamine - a compound known to increase cytoskeletal prestress but not directly influence fibroblast to myofibroblast transition (Cat# H7125, Sigma-Aldrich), for 72 hours (39). Our observations from N=3 independent trials reveal that 16.6±4.5 fibroblasts in the fibroblast mono-culture transitioned into myofibroblasts as compared to only 4.3±1.5 fibroblasts in untreated controls, suggesting that an increase in cytoskeletal prestress alone can induce fibroblast to myofibroblast transition (**Fig. 4A-D,** unpaired t-test, p=0.011, N=3). The temporal relationship between the increase in cytoskeletal tension and the subsequent phenotypic transition suggests a causal link between these two phenomena. These findings not only shed light on the mechanical aspects of the fibroblast-to-myofibroblast transformation but also underscore the importance of cell-cell interactions and mechanical forces in regulating cellular behavior and FMT. This contributes to our understanding of the complex interplay between cellular mechanics and phenotypic plasticity in the context of tissue remodeling and fibrosis.

### 5. Mechanical activation of GqGPCR pathway is essential for contact-based FMT

It is well-established that signal transduction across cell membranes involves the activation of two key enzymes: phosphatidylinositol (PI)-specific phospholipase C (PLC) and phosphoinositide 3-kinase (PI 3-kinase)(40). Typically, PLC activation occurs in response to various cellular agonists. In our study, we hypothesized that cells in contact utilize mechanical force as a stimulus to initiate PLC activation (**Fig. 5A**). Specifically, we hypothesized that a shear stress of 20Pa is sufficient to trigger PLC activation and Calcium release ($Ca^{2+}$). Myofibroblasts can exert shear stress of the order of 1kPa(41). The shear stress that we apply here is a small fraction of the stress exerted by a myofibroblast. This shear generates force to activate Phospholipase C, triggering the hydrolysis of Phosphatidylinositol 4,5-bisphosphate (PI(4,5)P2 or PIP2). The hydrolysis results in the production of two important second messengers: inositol 1,4,5-trisphosphate (I(1,4,5)P3 or IP3) and diacylglycerol (DAG)(40,42,43). IP3 is released from the plasma membrane and diffuses to IP3 receptors (IP3Rs) on the endoplasmic reticulum (ER) further releasing $Ca^{2+}$ which in turn promotes actomyosin cycling and cellular force generation.(44)

Continuous time-lapse imaging of the cell ensemble was performed to record the fluorescence intensities of PIP2 or $Ca^{2+}$ for 5-minutes (baseline). Next, to test the existence of a force-induced PLC activation mechanism in fibroblasts, we used an indenter to apply a shear stress of 20Pa calculated based on displacements of beads attached to the gel, and the stiffness of the gel. We observed a systematic decline in PIP2 levels and increase in calcium [$Ca^{2+}$] levels (measured as fluorescence intensity) in the cell ensemble in comparison to baseline fluorescence intensity. Confirming PLC activation, PIP2 hydrolysis and $Ca^{2+}$ release (**Fig. 5B**). To test whether mechanical activation of Gαq pathways was involved in contact-based FMT transitions, we used FR900359 (Cat# 33666, Cayman Chemical) a cyclic depsipeptide inhibitor of Gαq, Gα11, and Gα14 in a co-culture of fibroblasts and myofibroblasts on soft (E=300Pa) substrate(45). Cells were grown under standard culture conditions and treated with 10μM FR900359 for 72 hours (controls were samples that did not receive inhibitor). Staining for myofibroblast markers revealed that when the co-culture was treated with 10μM FR900359 only $2.33 \pm 0.57$ fibroblasts transitioned into myofibroblasts in the presence of the GqGPCR inhibitor despite being in contact with myofibroblasts as compared to $22 \pm 3.5$ cells in untreated co-cultures controls (**Fig. 5D-E,** $p < 0.001$, unpaired t-test, N=3). Myofibroblasts remained unaffected under both conditions. These findings provide valuable insight into the molecular mechanism underlying fibroblast to myofibroblast transitions, suggesting that the GqGPCR pathway is triggered upon myofibroblast contact with fibroblast and inhibition of this pathway prevents cell-cell contact-induced transition.

## 6. Endogenous TGF-β inhibition reveals a tension-dependent maintenance mechanism in myofibroblasts

To this point in our investigations, the role of endogenous TGF-β activation was not explicitly considered(46–49). To assess the role of TGF-β, we tested whether blocking TGF-β signaling alters contact-mediated phenotype change. Fibroblast-myofibroblast co-cultures were plated onto 300Pa substrates and treated with 10µM TGF-β receptor inhibitor, SB431542 (0.1% DMSO) for 72 hours (Cat# 1614, Tocris Bioscience). Cells were then stained for myofibroblast phenotype and revealed no fibroblasts transitioned despite being in a co-culture as compared to 20±5 cells in untreated co-cultures (**Fig. 6.1,** p = 0.0215, unpaired t-test, N=3), suggesting the involvement of endogenous TGF-β signaling in FMT. However, the mechanism by which TGF-β facilitates transition in this system remains unclear. We hypothesized that this effect is mediated through regulation of cytoskeletal tension.

To test our hypothesis , we cultured myofibroblasts on soft (E = 1.9kPa) beaded substrates for 24 hours. The following day, we performed continuous live-cell imaging to measure traction forces in myofibroblasts on soft substrates before and after TGF-β inhibition. Simultaneously cytosolic $Ca^{2+}$ dynamics were also monitored using a fluorescent $Ca^{2+}$ indicator. Following 5-minutes of baseline traction and $Ca^{2+}$ acquisition (no inhibitor), cells were treated with 10µM SB431542 (in 0.1% DMSO) or only (0.1% DMSO, controls). Interestingly, there was an immediate and statistically significant reduction in cellular traction forces within 15-minutes of treatment in comparison to baseline traction forces (p=0.0042, paired t-test, N=3) and DMSO controls (p=0.0348, unpaired t-test, N=3). Whereas 0.1% DMSO-treated controls maintained baseline traction levels (**Fig. 6.2,** ns, unpaired t-test, N=3). This indicates that although endogenous TGF-β may contribute to contact-mediated FMT, its functional impact is channeled through rapid maintenance of cytoskeletal tension. What was also particularly notable was that inhibitor treatment triggered rapid, sustained loss of cytosolic calcium, with no recovery over the 15-minute imaging window (see **Supplemental Figure S3**). This suggests that inhibitor treatment perturbs $Ca^{2+}$ homeostasis/influx mechanisms, dampening $Ca^{2+}$ dependent contractility, suppressing contact-mediated phenotype change. Together, these results indicate that the role of endogenous TGF-β in our system enabling fibroblast transition is tightly coupled to mechanical maintenance of cellular pre-stress. Thereby supporting our central premise that cytoskeletal tension rather than TGF-β secretion alone, governs FMT.

## DISCUSSION

This study uncovers novel mechanisms driving fibroblast-to-myofibroblast transition (FMT) in fibrosis, challenging the traditional view that FMT is primarily governed by soluble factors or a stiff extracellular matrix (ECM). We demonstrate that direct cell-cell contact between fibroblasts and myofibroblasts is sufficient to induce FMT, independent of the known exogenous biochemical signals or stiff matrix. Our findings emphasize the critical role of intercellular mechanical signaling and highlight the GqGPCR pathway as a key mediator of contact-induced FMT. We demonstrate that myofibroblast contact with a fibroblast initiates GqGPCR activation in the fibroblast and drives cytosolic $Ca^{2+}$ release. This leads to actomyosin contractility and cellular force generation, promoting the emergence of the myofibroblast phenotype. Importantly, we also show that inhibition of Gαq/11/14 can prevent FMT, presenting a promising target for anti-fibrotic therapies. Notably, the observed reduction in cellular traction following inhibition of endogenous TGF-β was not anticipated, but it highlighted an unappreciated coupling between prestress-dependent mechanics and endogenous TGF-β linked responses in our system. TGF-β inhibition also triggered a sustained loss of cytosolic $Ca^{2+}$, a key mediator of contractility in our pathway. When $Ca^{2+}$ is depleted, traction forces rapidly decline and the transition is suppressed, reinforcing cytoskeletal tension as a key regulator of contact-mediated FMT. Furthermore, the observed increase in cytoskeletal tension preceding a phenotype change and driving FMT suggests that modulating cellular mechanics could be an effective therapeutic strategy.

We used an in vitro, linearly elastic soft-substrate system to avoid confounding matrix nonlinearities of fibrillar collagen(50,51). Our platform enables precise control of material properties of the substrate, allowing quantitative assessment of contact-driven fibroblast-myofibroblast transitions. An important next step will be to determine the extent of fibrotic activation in our system, by quantifying cell-type specific ECM deposition within the co-culture. It will also be useful to further identify specific adhesion molecules (e.g. cadherins, integrins, nectins, or ephrin/Eph receptors) responsible for sensing contact and activating the GqGPCR pathway. While these analyses are beyond the scope of the present study, it is necessary for linking our mechanistic findings to functional matrix remodeling outcomes.

Fibrosis is not a single disease, but a shared pathological outcome of many chronic injuries - including fibrotic airway remodeling in asthma, idiopathic pulmonary fibrosis, liver cirrhosis, and cardiac fibrosis after infarction(12,52,53). The myofibroblast phenotype is central to fibrosis, yet its definition and persistence remain debated. We recognize the difficulty in defining the myofibroblast phenotype, however, the phenotype of a cell is not an intrinsic property of the cell. Rather, the cell phenotype is a set of emergent properties that result from cells interacting with other cells and their extracellular environment, including mechanical stretch in organs like the heart and the lung. In this study, we have used a classical marker of α-smooth muscle actin in the stress fibers. We found this marker to be consistent with the increase in cytoskeletal tension typical of myofibroblasts. We extend this literature by proposing a novel mechanism that explains why myofibroblasts persist even when inflammation is controlled, and the fibrotic matrix is removed. By highlighting the interplay between intercellular mechanical forces and cytoskeletal tension, this work opens new avenues for therapeutic interventions targeting the mechanical aspects of fibrotic disease. In conclusion, this study reveals a previously unrecognized mechanism of FMT driven by direct cell-cell interactions and mechanical signaling, expanding our understanding of fibrosis progression.

## SUPPLEMENTAL MATERIAL

## ACKNOWLEDGMENTS

The authors thank Ms. Christina Velez for providing the EFS-tdTomato lentivirus. The authors also thank Mr. Barrett Smith and Dr. Sara M Hashmi, Assistant Professor, Department of Chemical Engineering, Northeastern University for assistance and providing us with the Rheometer equipment. Preprint available at https://arxiv.org/abs/2503.01834.

## GRANTS

This work was supported by the National Science Foundation (NSF) Career grant #2047207 awarded to H.P.

## AUTHOR CONTRIBUTIONS

H.P. conceived the research and developed the overall hypothesis. V.C. and H.P. designed the experiments. V.C., B.M.G., and H.P. performed the experiments, analyzed the experimental data, and interpreted the results. V.C., and B.M.G. performed the monolayer stress traction, and traction force microscopy experiments, analyzed the data, prepared the figures, and interpreted the results. V.C. drafted the manuscript. V.C., B.M.G., and H.P. edited, and revised the manuscript. H.P. approved the final version of the manuscript.

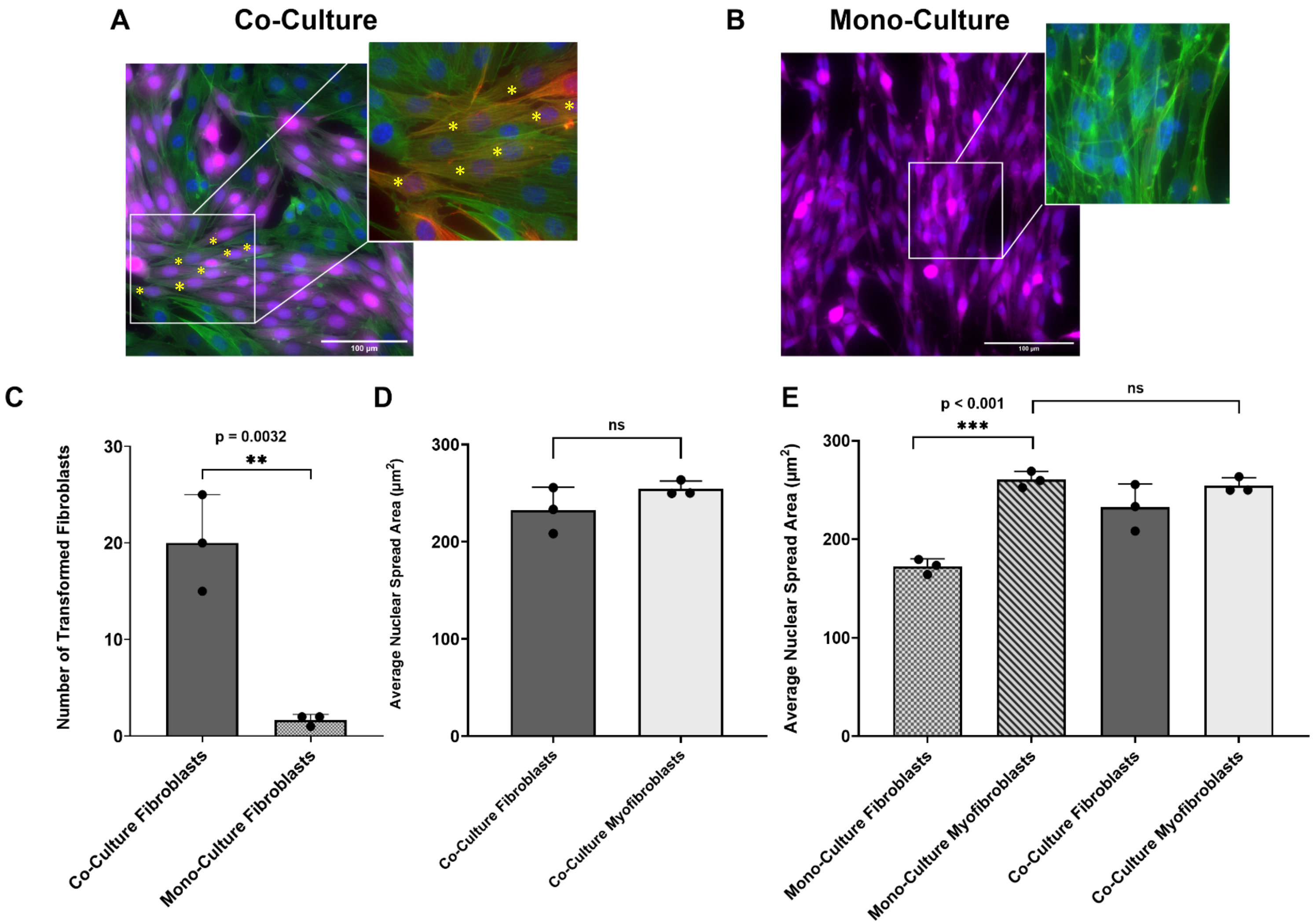

**Fig 1**: **Fibroblasts when co-cultured with myofibroblasts on a soft (E=300Pa) PDMS substrate, undergo a phenotypic shift to myofibroblasts despite the absence of external biochemical or stiff matrix stimuli**. Cells were stained for myofibroblast markers f-actin (green) and α-SMA (red) and nuclear stain DAPI (blue) after 3 days in culture. **(A)** Fibroblasts (magenta) in co-culture with myofibroblasts showed sequestration of α-SMA in their stress fibers when cultured on a 300Pa soft substrate, as in the inset (indicated by reddish-yellow stress fibers). The transitioned cells are indicated by yellow asterisks. **(B)** Fibroblasts in monoculture on 300Pa soft substrate retained their phenotype when not in contact with myofibroblasts. **(C)** A significant number of fibroblasts in co-culture (20±5 cells) transformed into myofibroblasts as compared to fibroblasts in monoculture (1.66 ± 0.57, $p = 0.0032$, unpaired t-test, $\eta^2 = 0.91$, N=3). **(D)** There is no significant difference in the nuclear spread area of fibroblasts in a co-culture compared to myofibroblasts in the co-culture (ns, paired t-test, N=3). **(E)** There is a significant difference in the nuclear spread area of fibroblasts in a mono-culture when compared to myofibroblasts in a mono-culture on a soft substrate (172 ± 7.66 $\mu m^2$, $p<0.001$, unpaired t-test, $\eta^2 = 0.97$, N=3). However, there is no significant difference in the nuclear spread area of fibroblasts and myofibroblasts in the co-culture when compared to that of myofibroblasts in a monoculture (One-way ANOVA, ns, N=3). All data represented as mean±SD. Scale bars 100µm.

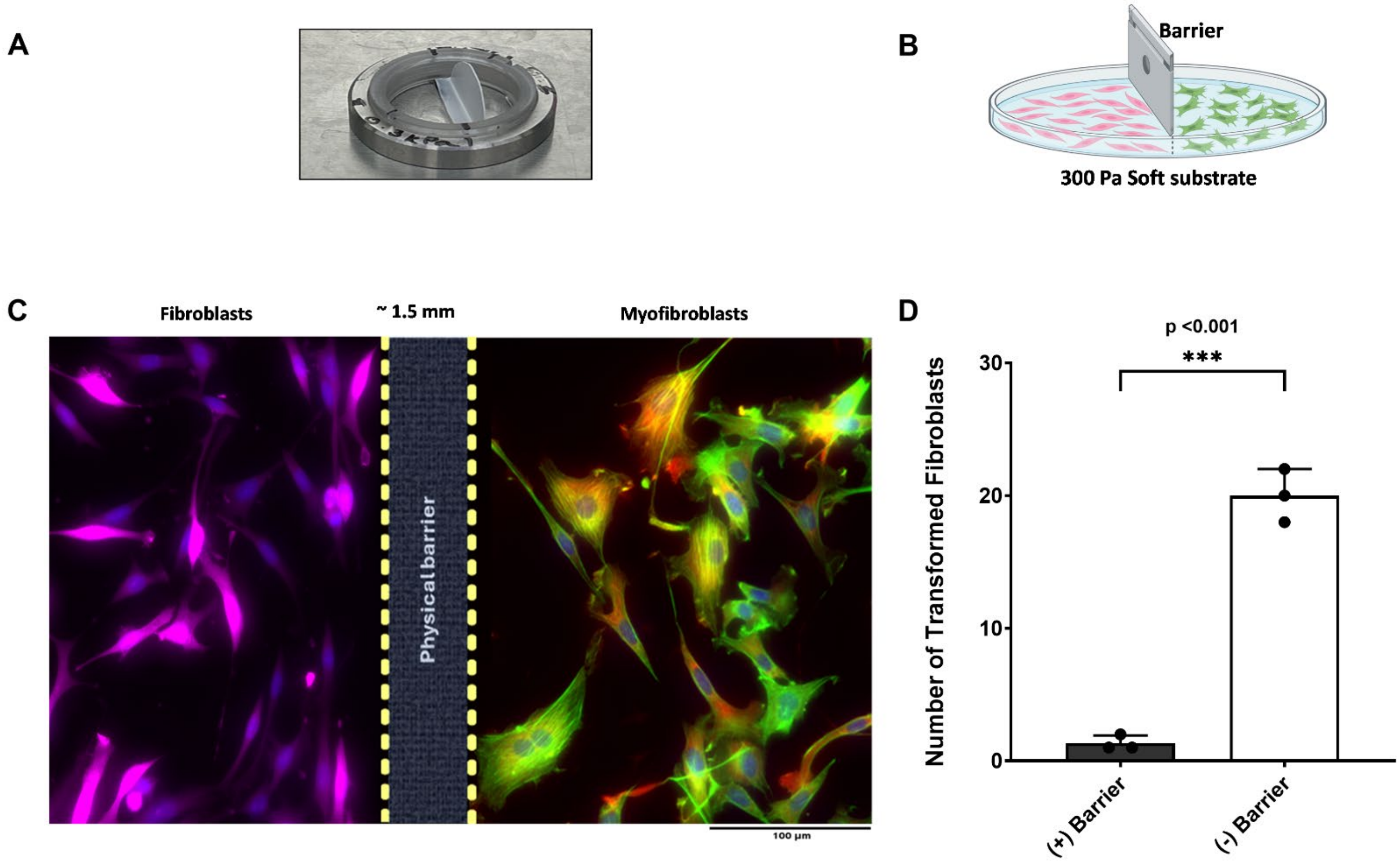

**Fig 2: Cell-cell contact is sufficient for fibroblast-to-myofibroblast transitions. (A)** A picture of the physical barrier set up on the 300Pa soft substrate in a bioptech dish to separate fibroblasts from the myofibroblasts. **(B)** Diagrammatic representation of experimental setup shows fibroblasts in pink and myofibroblasts in green on either side of the barrier. **(C)** Cells were stained for myofibroblast markers, F-actin (green) and α-SMA (red), and nuclear stain DAPI (blue) after 3 days in culture. Fibroblasts (magenta) did not express α-SMA in their stress fibers when not in contact with myofibroblasts despite the presence of intercellular chemical signals, but myofibroblasts retained their phenotype (visible reddish-yellow stress fibers). **(D)** With the barrier in place, only 1.3$\pm$0.57 fibroblasts showed signs of alpha-smooth muscle actin in stress fibers and in the absence of divider, 20$\pm$2 fibroblasts transitioned into myofibroblasts ($p < 0.001$, unpaired t-test, $\eta^2 = 0.99$, N=3). Data represented as mean$\pm$SD. Scale bar 100µm.

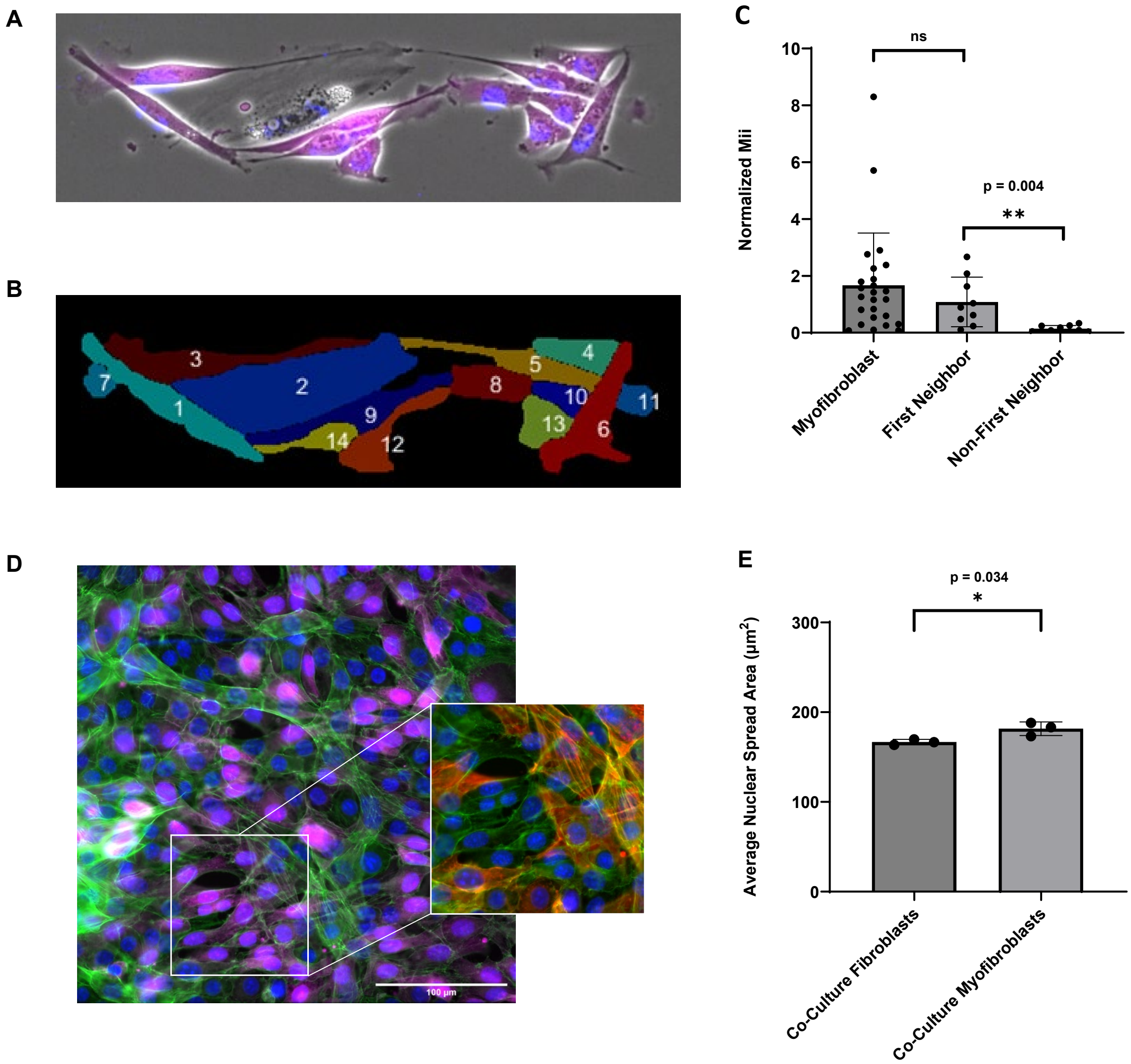

**Fig 3: Fibroblasts in contact with myofibroblasts exhibit increased cytoskeletal tension in 24 hours (1 day of contact with myofibroblasts). (A)** Representative image of a patterned cellular ensemble of a myofibroblast-fibroblast co-culture. Fibroblasts are shown in magenta, and myofibroblast is distinguishable as the non-magenta cell. Nuclei of both cell types are counterstained with DAPI (blue). Scale bar 50µm **(B)** Here, cell 2 is a myofibroblast, with 1, 3, 5 and 9 considered "first neighbor fibroblast" as they are in direct contact with the myofibroblast, and the remaining cells are "non-first neighbor fibroblasts" as they are not in direct contact with myofibroblasts. **(C)** Fibroblasts in contact with a myofibroblast exhibit increased contractility ($M_{ii}$) within 24 hours, reaching levels comparable to myofibroblasts, indicating an early mechanical transition before a documented phenotypic change (ns, Mann-Whitney Rank Sum Test, N=2). In contrast, fibroblasts without direct myofibroblast contact show significantly lower contractility ($p = 0.004$, Mann-Whitney Rank Sum Test, N=2). **(D)** Fibroblasts do not express α-SMA in their stress fibers, indicating that they do not transition into myofibroblasts despite being

in contact with myofibroblasts for 24 hours (indicated by the absence of reddish-yellow stress fibers). Scale bar 100µm. **(E)** The average nuclear spread area of fibroblasts in the co-culture, is significantly lower than that of the myofibroblasts in the same culture, indicating a lack of phenotype transition ($p=0.034$, One-way ANOVA, $\eta^2 = 0.91$, $N=3$).

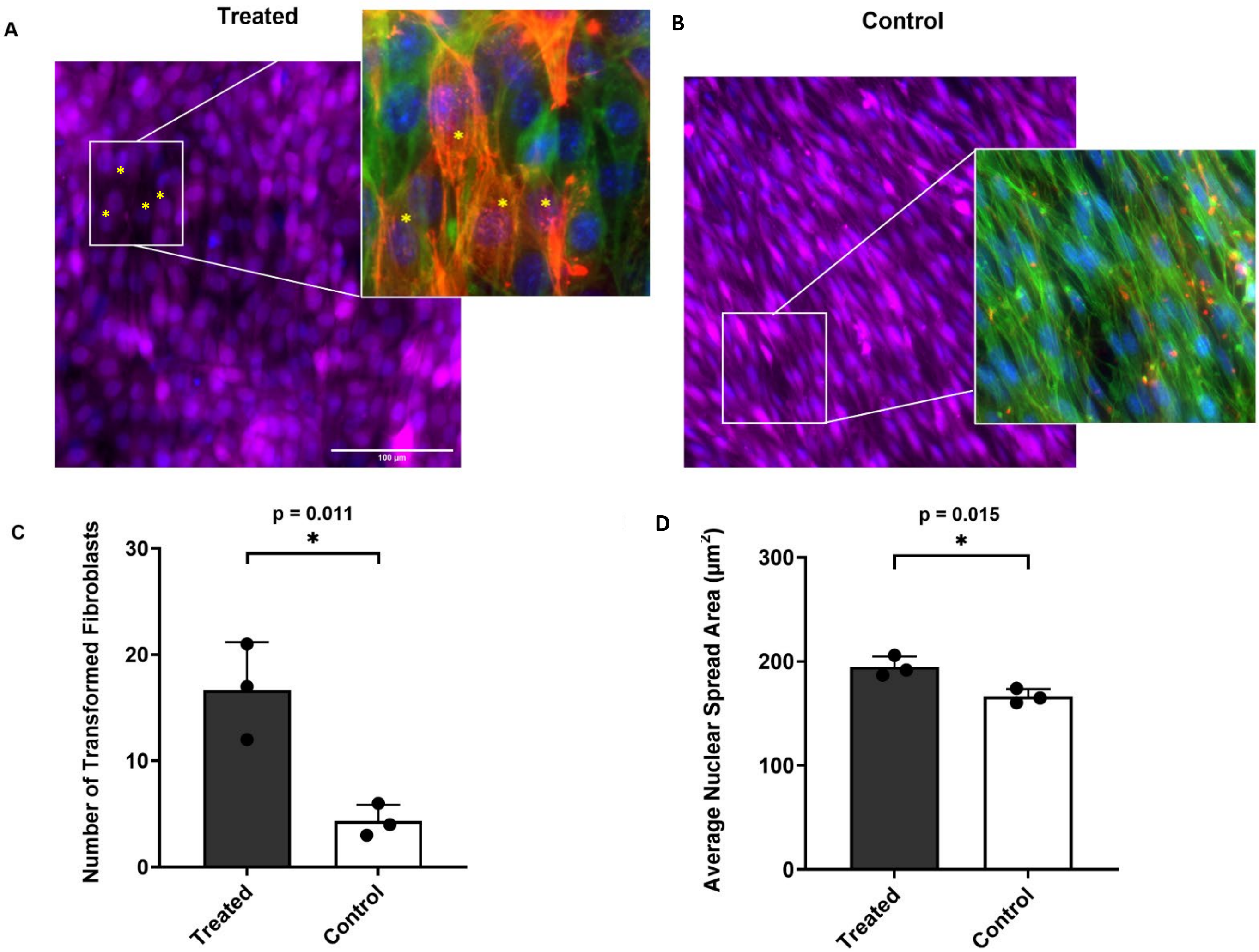

**Fig 4: A rise in cytoskeletal pre-stress is sufficient to induce the transition of fibroblasts to myofibroblasts even in the absence of contact with a myofibroblast on a 300Pa soft substrate. (A)** Fibroblast mono-cultures (magenta) treated with 2.5µM histamine (72 hours) to increase their cytoskeletal pre-stress show sequestration of α-SMA in their stress fibers (indicated by reddish yellow stress fibers). The transitioned cells are indicated by yellow asterisks. **(B)** Fibroblast mono-culture controls (untreated) do not show α-SMA in their stress fibers (absence of reddish yellow stress fibers). **(C)** A significant number of fibroblasts (16.6 ± 4.5 cells) transformed into myofibroblasts in response to increased cellular pre-stress alone when on a 300Pa soft substrate, compared to the controls, 4.3 ± 1.5 cells (p=0.011, unpaired t-test, $\eta^2$ = 0.83, N=3). **(D)** The average nuclear spread area of fibroblasts in histamine treated co-cultures is significantly higher (194.8 ± 9.8 $\mu m^2$) than controls (166 ± 7.03 $\mu m^2$), indicating a phenotype transition due to a rise in cellular pre-stress (p=0.015, unpaired t-test, $\eta^2$ = 0.82, N=3). Data represented as mean±SD. Scale bars 100µm.

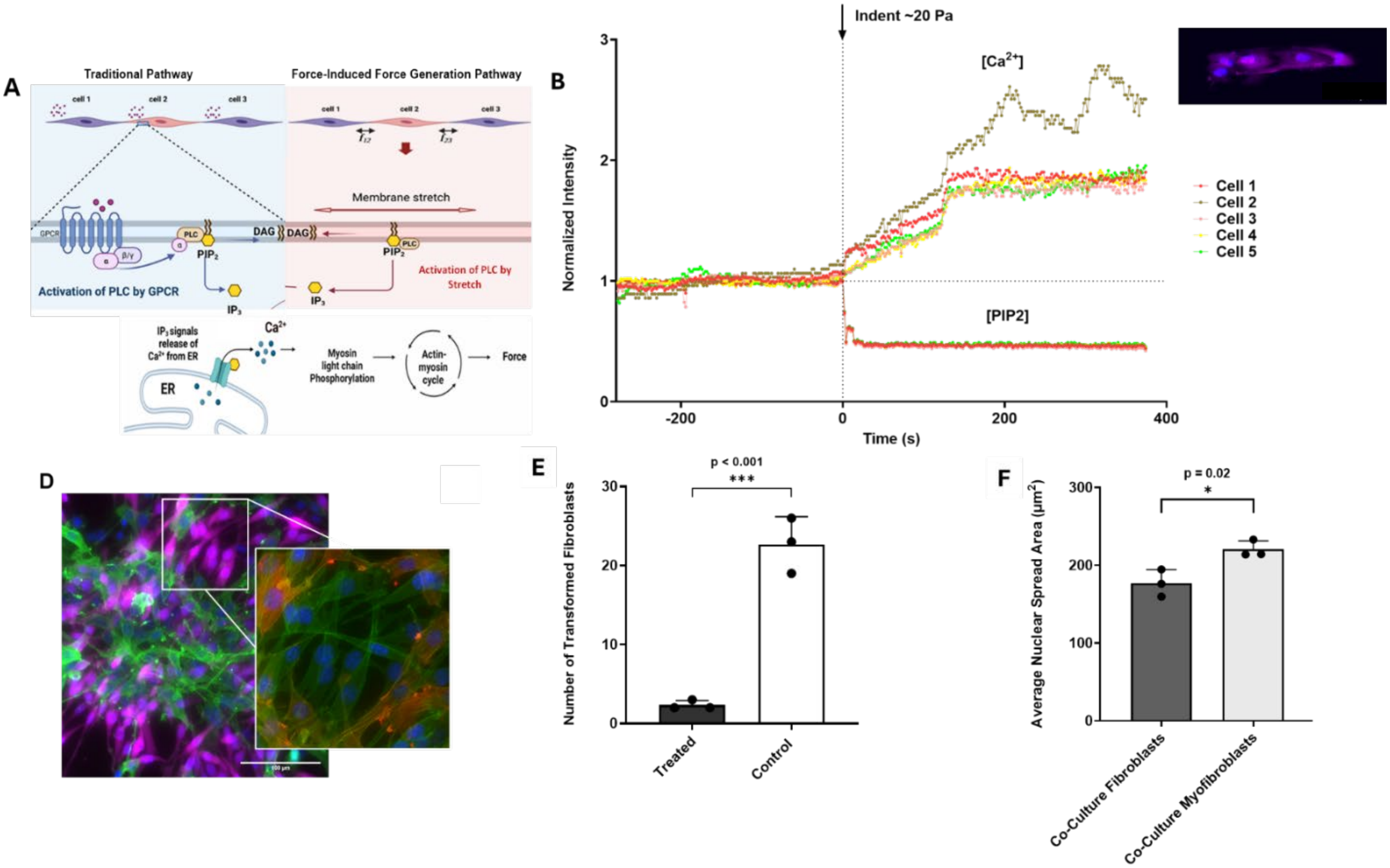

**Fig 5: Cell-cell contact induces PIP2 hydrolysis and blocking the GqGPCR pathway prevents cell-contact induced fibroblast to myofibroblast transitions on a 300Pa soft substrate. (A)** The diagram shows a force-induced force-generation mechanism in fibroblasts. Here, cells in contact utilize mechanical force as a stimulus to initiate PLC activation, just as an agonist would in the traditional pathway. **(B)** A shear stress of ~20 Pa was sufficient to induce PLC activation and result in PIP2 hydrolysis and Calcium release, indicated by a steady decline in PIP2 rise in calcium fluorescence intensity (N=3). The inset shows a micropatterned island of tdTomato labelled fibroblasts (Magenta) with nuclei counterstained with DAPI (blue). **(C)** When cells were stained for myofibroblast markers, F-actin (green) and α-SMA (red) and nuclear stain D API ( blue), fibroblasts do not show α-SMA in their stress fibers, indicating that they do not transition into myofibroblasts in the presence of the GqGPCR inhibitor despite being in a co-culture for 3 days. **(D)** Only a negligible fraction of fibroblasts (2.33 ± 0.57 cells) transitioned to myofibroblasts in the presence of the inhibitor as compared to 22 ± 3.5 cells in untreated co-cultures ($p < 0.001$, unpaired t-test, $\eta^2 = 0.96$, N=3, data represented as mean±SD). **(E)** The average nuclear spread area of fibroblasts in the co-culture when GqGPCR pathway is inhibited, is significantly lower than that of the myofibroblasts in the same culture, indicating a lack of phenotype transition ($p = 0.02$, paired t-test, $\eta p^2 = 0.96$, N=3). Scale bar 50µm.

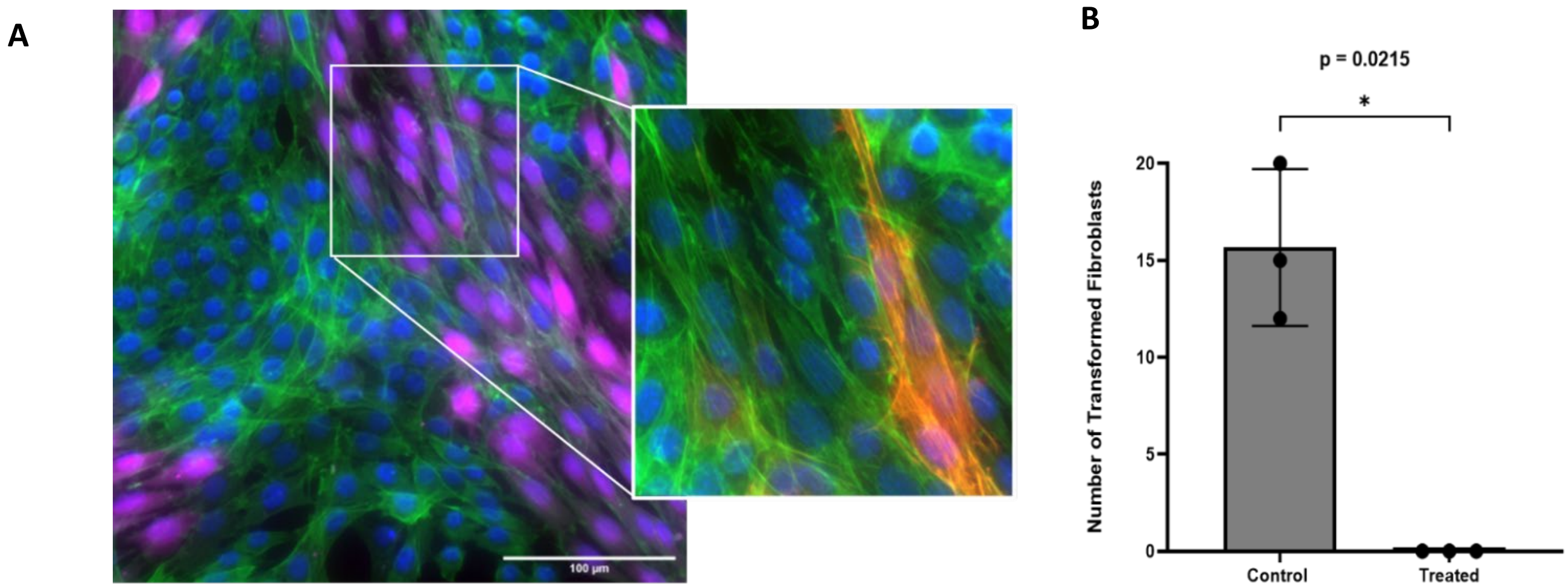

**Fig 6.1. A coculture treated with 10 µM SB431542 TGF-β inhibitor over three days does not show any transition in fibroblasts. (A)** Fibroblasts (magenta: tdTomato, nuclei stain: DAPI) in co-culture with myofibroblasts (green: f-actin, nuclei stain: DAPI) showed no sequestration of α-SMA in their stress fibers when cultured on a 300Pa soft substrate. As seen in the inset, fibroblasts displayed no reddish-yellow stress fibers (green: f-actin, nuclei stain: DAPI), whereas myofibroblasts retained their phenotype indicated by α-SMA+ stress fibers (reddish yellow stress fibers, nuclei stain: DAPI). **(B)** A significant number of fibroblasts (20±5 cells) transformed into myofibroblasts in untreated samples as compared to no transition of fibroblasts in cocultures treated with the inhibitor ($p = 0.0215$, $\eta^2 = 0.96$, unpaired t-test, N=3, data represented as mean ± SD). Scale bar 100µm.

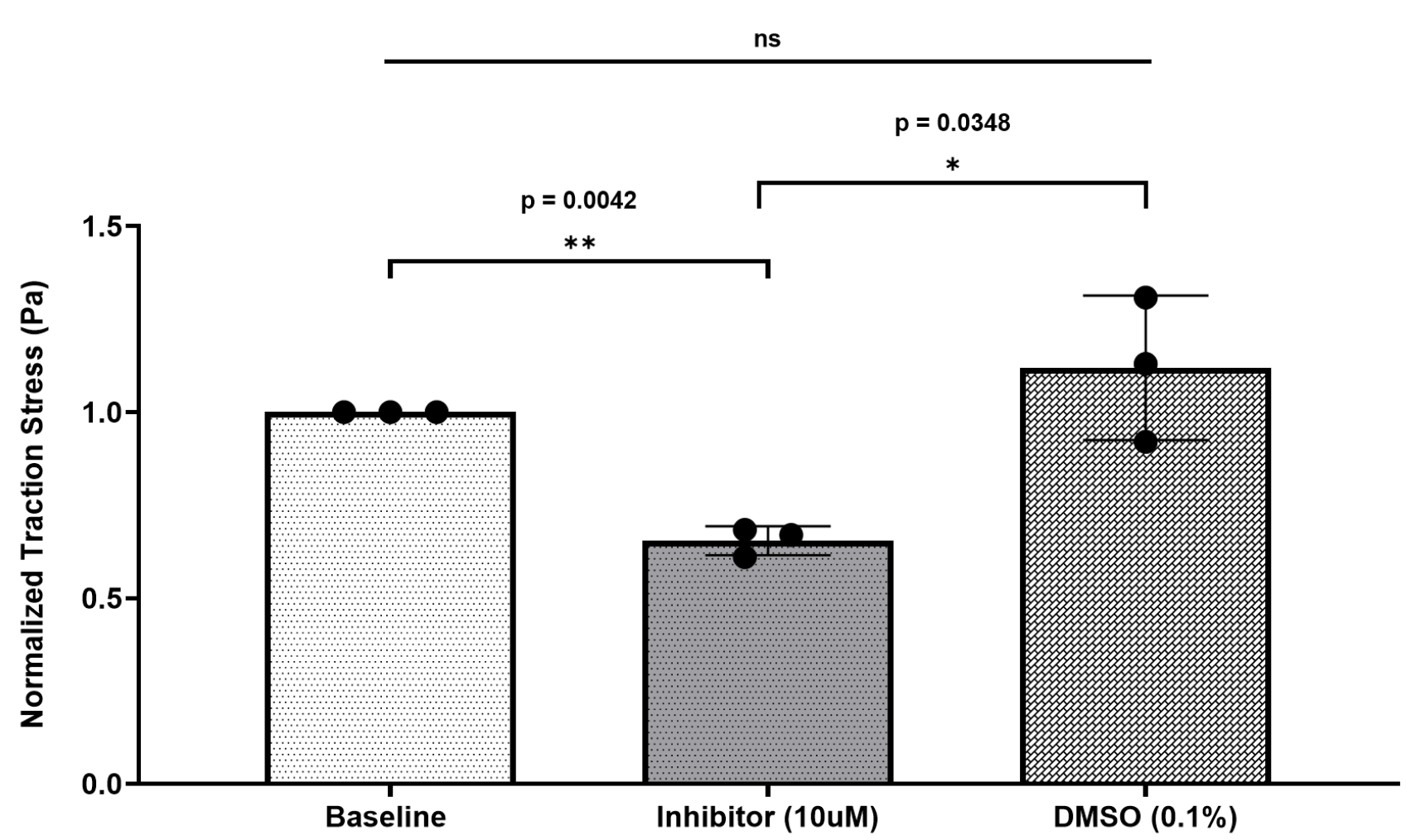

**Fig 6.2. Early reduction in traction forces following TGF-β inhibition on soft substrates.** Myofibroblasts cultured on 300Pa soft NuSil substrates were treated with 10 µM SB431542 (0.1% DMSO) or 0.1% DMSO (control). Within 15-minutes of inhibitor treatment, traction stresses declined significantly in comparison to baseline tractions (paired t-test,

p=0.0042, 0.6544 ± 0.034, N=3) indicating a rapid loss of cytoskeletal tension. In contrast, DMSO treated cells maintained stress levels following treatment in comparison to baseline tractions (ns, paired t-test, N=3). Direct comparison between inhibitor-treated and control cells confirmed a significant difference in traction forces (unpaired t-test, p=0.0348, 0.836 ± 0.345, $\eta^2$ = 0.81, N=3). Thereby demonstrating that TGF-β inhibition significantly reduces cellular traction, underscoring the importance of cytoskeletal stress as a critical and early regulator of phenotypic transitions. All data are represented as mean ± SD.

## Supplemental Figure S1:

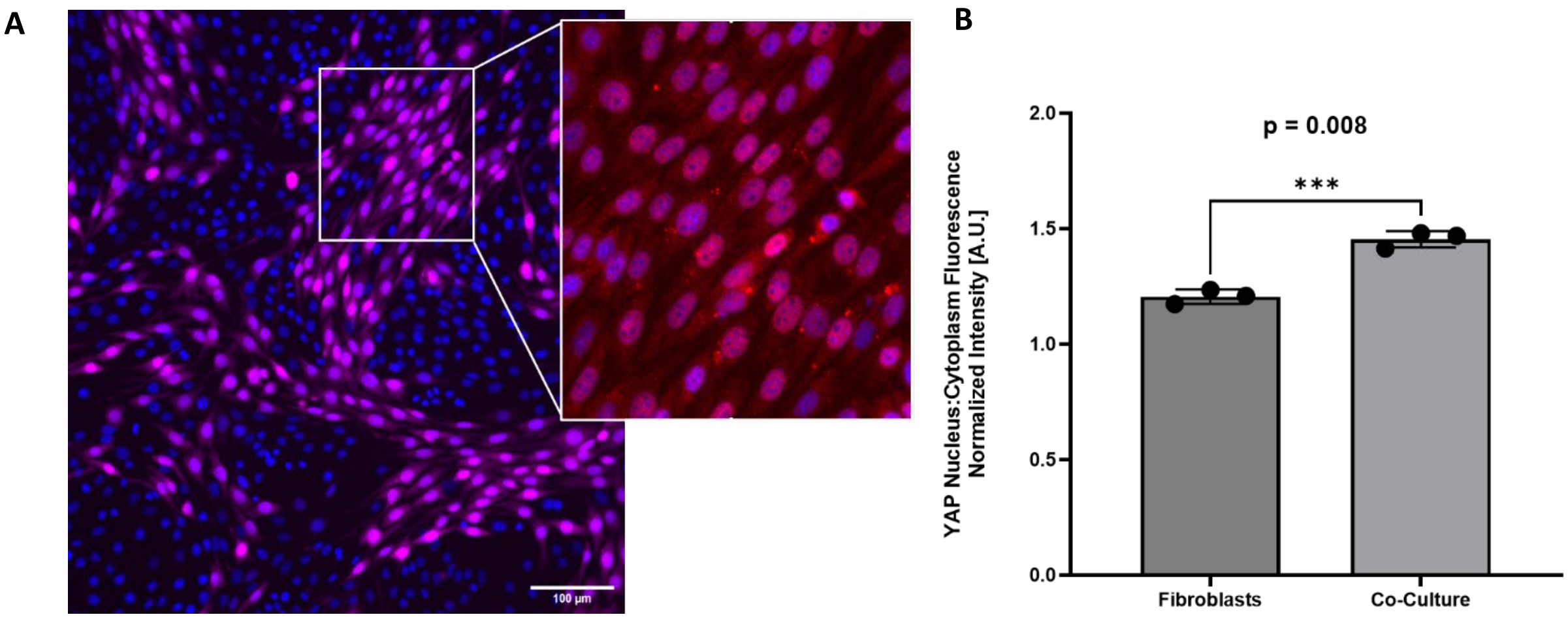

**Supplemental Figure S1. Fibroblast contact with myofibroblasts induces significant YAP nuclear localization in fibroblasts**. **(A)** Representative 20X image of Fibroblasts (magenta) in direct contact with myofibroblasts (unlabeled, DAPI-stained) display predominantly nuclear localization of YAP (red). The inset highlights nuclei appearing red due to YAP accumulation, indicating its translocation into the nucleus. In contrast, fibroblasts cultured alone on a soft substrate exhibit predominantly cytoplasmic YAP localization, with YAP either distributed between the cytoplasm and nucleus or largely excluded from the nucleus. **(B)** Quantification of the nuclear-to-cytoplasmic (N:C) fluorescence intensity ratio shows a significant increase in co-cultured fibroblasts compared to monoculture, confirming enhanced nuclear accumulation of YAP upon myofibroblast contact. (p = 0.008, Welch's unpaired t-test, N=3). Scale bar 100µm.

## Supplemental Figure S2:

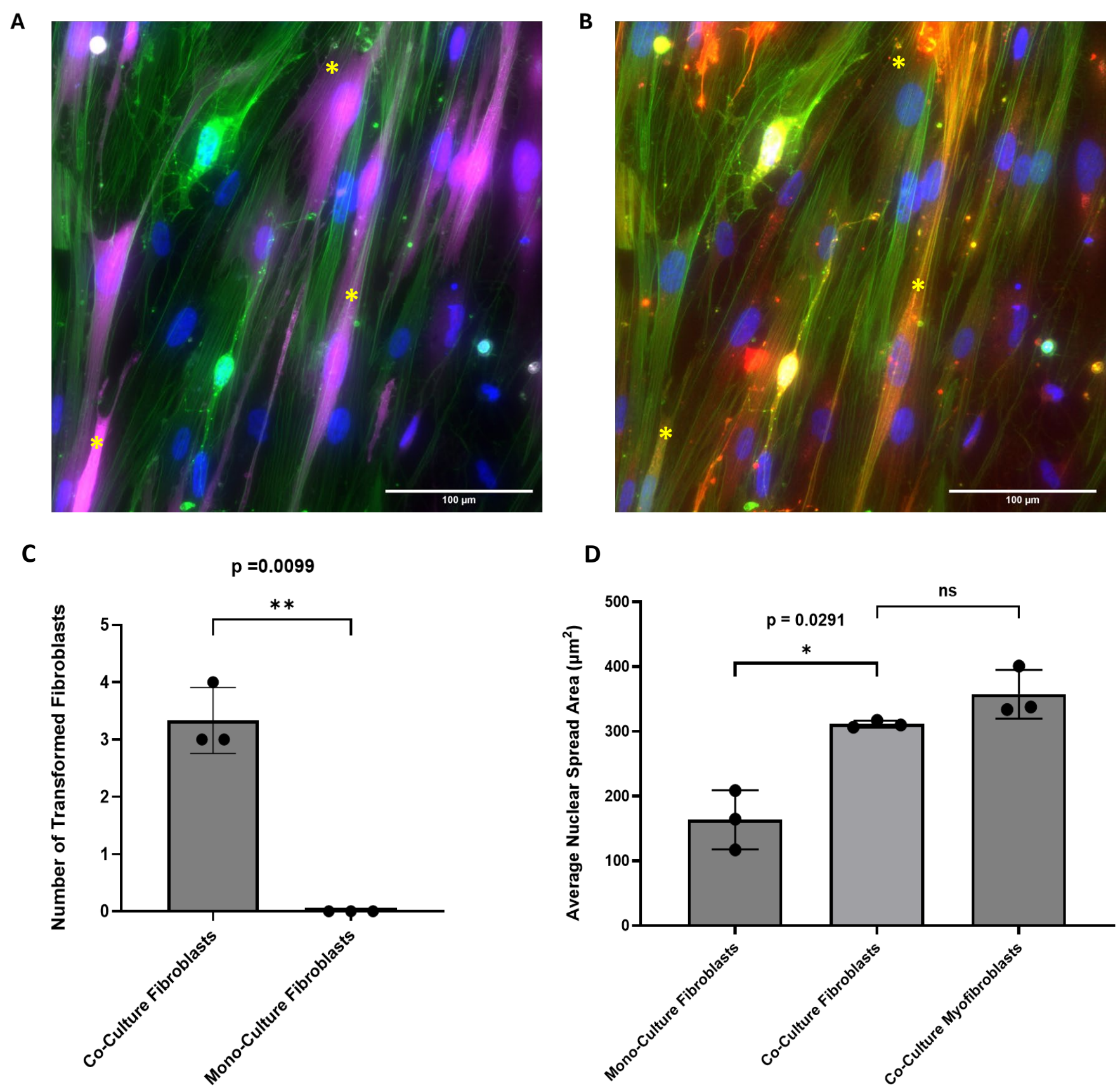

**Supplemental Figure S2. Primary healthy normal human lung Fibroblasts when co-cultured with normal human lung myofibroblasts on a soft (E=300Pa) PDMS substrate, undergo a phenotypic shift to myofibroblasts.** Cells were stained for myofibroblast markers f-actin (green) and α-SMA (red) and nuclear stain DAPI (blue) after 3 days in culture. **(A)** Representative fibroblasts (magenta, DAPI) in co-culture with myofibroblasts (f-actin, DAPI). **(B)** After 3 days in culture with myofibroblasts, fibroblasts showed sequestration of α-SMA in their stress fibers (indicated by reddish-yellow stress fibers). The transitioned cells are indicated by yellow asterisks. **(C)** A statistically significant number of fibroblasts (3.33 ± 0.58 cells) transitioned to myofibroblasts compared to no transitions observed in fibroblast mono-cultures ($p = 0.0099$, $\eta^2 = 0.98$, unpaired t-test, N=3). **(D)** There is no statistically significant difference in the nuclear spread area of fibroblasts in co-culture compared to myofibroblasts in the co-culture (ns, paired t-test, N=3). However, there is a statistically significant difference in the nuclear spread area when fibroblasts in coculture are compared to fibroblasts in mono-culture ($p = 0.0291$, $\eta^2 = 0.94$, unpaired t-test, N=3). All data are represented as mean ± SD per 40X field of view. Scale bars 100µm.

## Supplemental Figure S3:

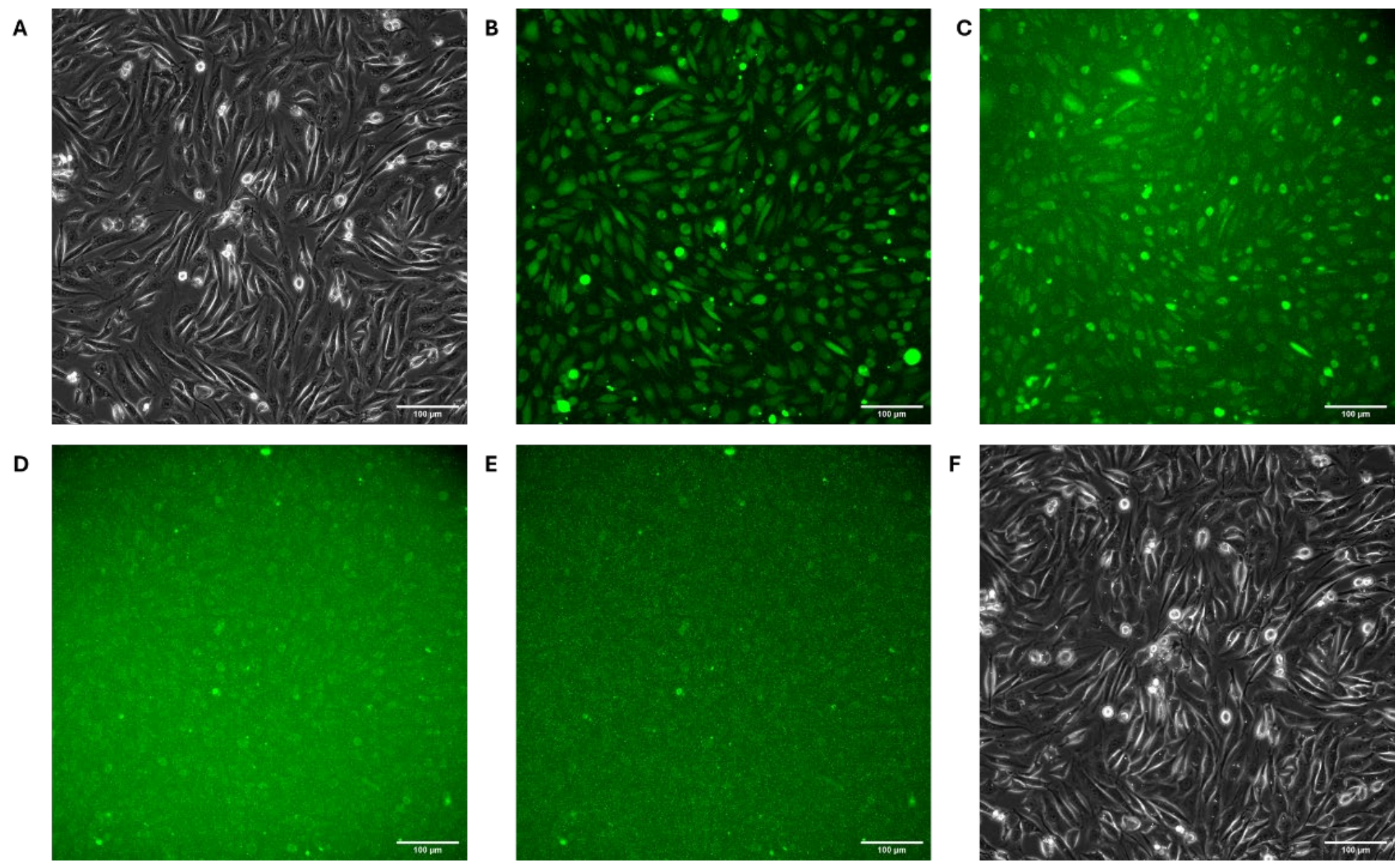

**Supplemental Figure S3. TGF-β Inhibitor induces rapid and sustained loss of intracellular $Ca^{2+}$ signal in myofibroblasts.** Representative 20X fluorescence images of $Ca^{2+}$ labeled myofibroblasts acquired consecutively alongside traction force measurements. **(A)** Baseline phase image prior to inhibitor addition. **(B)** Baseline intracellular $Ca^{2+}$ fluorescence prior to inhibitor addition. **(C)** Five minutes post-treatment, intracellular $Ca^{2+}$ signal begins to diminish, consistent with early $Ca^{2+}$ depletion. **(D)** At 10-minutes post-treatment, intracellular $Ca^{2+}$ fluorescence is markedly reduced to near-background levels. An apparent increase in surrounding extracellular fluorescence is observable, suggesting robust $Ca^{2+}$ efflux into the medium. **(E)** At 15-minutes post-treatment, intracellular $Ca^{2+}$ signal does not recover within the imaging window. **(F)** Post-treatment phase image shows cells are viable, with no observable signs of cell death. Together with the observed drop in traction forces, this data indicates that the inhibitor disrupts myofibroblast contractility and is associated with acute perturbation of $Ca^{2+}$homeostasis. Scale bars 100µm.